\documentclass{pasj00}
\begin{document}
\SetRunningHead{K. Sugitani et al.}{Polarimetry in M16}
\Received{2007/01/12}%{yyyy/mm/dd}
\Accepted{2007/03/19}%{yyyy/mm/dd}

\title{Near-Infrared Polarimetry of the Eagle Nebula (M16)}

%%% begin:list of authors
 \author{%
Koji \textsc{Sugitani},\altaffilmark{1}
Makoto \textsc{Watanabe},\altaffilmark{2}
Motohide \textsc{Tamura},\altaffilmark{3}
Ryo \textsc{Kandori},\altaffilmark{3}
J. H. \textsc{Hough},\altaffilmark{4}\\
Shogo \textsc{Nishiyama},\altaffilmark{3}
Yasushi \textsc{Nakajima},\altaffilmark{3}
Nobuhiko \textsc{Kusakabe},\altaffilmark{5}
Jun \textsc{Hashimoto},\altaffilmark{1,6}\\
Takahiro \textsc{Nagayama},\altaffilmark{7}
Chie \textsc{Nagashima},\altaffilmark{8}
Daisuke \textsc{Kato},\altaffilmark{8}
and 
Naoya \textsc{Fukuda}\altaffilmark{9}
  }
\altaffiltext{1}{Graduate School of Natural Sciences, Nagoya City University, Mizuho-ku, Nagoya 467-8501}
\altaffiltext{2}{Subaru Telescope, National Astronomical Observatory of Japan, Hilo, HI 96720, USA}
\altaffiltext{3}{National Astronomical Observatory of Japan, 2-21-1 Osawa, Mitaka, Tokyo 181-8588}
\altaffiltext{4}{Center for Astrophysics Research, University of Hertfordshire, Hatfield, Herts AL10 9AB, UK}
\altaffiltext{5}{Graduate University of Advanced Science, 2-21-1 Osawa, Mitaka, Tokyo 181-8588}
\altaffiltext{6}{Department of Astrophysics, Tokyo University of Science, Shinjuku-ku, Tokyo 162-8601}
\altaffiltext{7}{Department of Astrophysics, Kyoto University, Sakyo-ku, Kyoto 606-8502}
\altaffiltext{8}{Department of Astrophysics, Nagoya University, Nagoya 464-8602}
\altaffiltext{9}{Department of Computer Simulation, Okayama University of Science, Okayama 700-0005}
%% `\KeyWords{}' always has to be placed before `\maketitle'.
 %Do NOT move this preamble from here!

\maketitle

\begin{abstract}
We carried out  deep and wide ($\sim$8\arcmin $\times$ 8\arcmin) $JHKs$ imaging
polarimetry in the southern region of the Eagle Nebula (M16).
The polarization intensity map reveals that two YSOs with near-IR reflection nebulae
are located at the tips of two famous molecular pillars (Pillars 1 and 2)
facing toward the exciting stars of M16.
The centrosymmetric polarization pattern are consistent with those around class I objects 
having circumstellar envelopes,
confirming that star formation is now taking place at the two tips
of the pillars under the influence of UV radiation from the exciting stars.
Polarization measurements of point sources show that magnetic fields are 
aligned along some of the pillars but in a direction that is quite different to the global 
structure in M16.

\end{abstract}

\section{Introduction}

Recent analyses of meteorites confirmed the presence
of Fe isotopes of supernova origin in the early solar nebula  and 
suggested that the solar system was formed in a massive star forming region 
\citep{ta03, ta06}.
\citet{he04} and \citet{he05} presented a formation scenario of the sun-like 
low-mass stars and their planetary systems in massive star forming regions. 
The M16 region is important as it is a site with such an environment.

The Eagle Nebula (M16) is one of the most noticeable star forming regions 
following a comprehensive study of the three molecular gas pillars (Pillars 1 -- 3) 
with HST \citep{he96}.  
Near-infrared studies revealed that the protostar-like or very young objects M16ES-1 
and M16ES-2
\citep{th02}, which were also identified as P1 and T1 by \citet{su02} and 
as YSO-1 and YSO-2 by \citet{mc02}, are located at the tips of Pillars 1 and 2, 
respectively, facing toward the main exciting star of M16.
\citet{su02} reported that the locations of these young stellar objects and 
the elongated structures of the pillars suggested their formation as being due to 
the interaction with UV light from the OB stars in the NGC 6611 cluster, 
in analogy with the sequential star formation in some bright-rimmed clouds \citep{su95}.
\citet{fu02} used interferometric observations of $^{13}$CO and C$^{18}$O ($J=1-0)$, 
and of 2.7 mm continuum to indicate the propagation of star formation activity in the heads 
of these two pillars due to radiation or wind from the main exciting star.
\citet{hea04} stated that the water masers in M16 were concentrated in the compressed gas layers 
within a few tenths of a parsec of the ionization front in three molecular pillars, 
one of which is Pillar 2, suggesting triggered star formation in their tip regions.  
\citet{he04} also discussed the sequential nature of star formation in the pillars.
For Pillar 3, \citet{th02} implied that the bright stars M16S-1 and M16S-2, which are located 
at the tip, were the results of recent star formation at the pillar head.

\citet{th02} estimated the luminosity of M16ES-1 and M16ES-2 to be 200 $L_{\odot}$ and 
20 $L_{\odot}$, respectively, as a single object or a cluster of objects. 
They suggested that M16ES-1 is not a source of ionization radiation 
based on the absence of Pa$\alpha$ emission, 
and that it is either made up of multiple lower-luminosity objects or
an object(s) in the earlier stage of ZAMS, i.e., protostar stage.
On the other hand, \citet{mc02} estimated the extinction of M16ES-1 and M16ES-2 
to be 27 and 15 mag., respectively, assuming their $Js$ and $H$ fluxes to be photospheric,  
and concluded that a 10 $M_{\odot}$ ZAMS star and 2.5 $M_{\odot}$ ZAMS star 
were located at the tips of Pillars 1 and 2, respectively.
It is not clear, from these observations, whether single stars of 
relatively high masses or multiple lower-mass stars have formed there.  
Thus, further observations are required to address this issue.

Recently, \citet{mi06} developed 3-D hydrodynamical model for the dynamical evolution 
of molecular clouds irradiated by UV light from massive stars, i.e., bright-rimmed clouds.  
They noted that since their model  adequately included self-gravity and 
treated the chemical and thermal evolution, it had advantage in the detailed 
study of triggered star formation over previous 2-D/3-D models that did not
fully include all of these effects (e.g., \cite{be89,le94,wh99, wi01, kb03}), although these models 
successfully reproduced some of the observed properties of bright-rimmed clouds.
\citet{mi06} applied their model for Pillar 2 and suggested that the core of Pillar 2 is 
at a transition stage toward induced star formation due to the shock that precedes 
an ionization front moving toward the core.  
In fact, a water-maser source, which indicates an object in the very early stage 
of stellar evolution, is detected near the core of Pillar 2 \citep{po98,wh99,fu02} 
a few arc-seconds inside of M16ES-2, which is located at the ionization front \citep{hea04}, 
suggesting further induced star formation in Pillar 2.   
However, these models do not predict magnetic field behavior in detail, 
although \citet{mi06} did include magnetic pressure and \citet{be89} used 
an approximate expression for the magnetic flux.
The lack of a full treatment of magnetic fields is due to the modeling complexity and 
to the paucity of observations of magnetic fields in and around molecular clouds 
associated with HII regions.

\citet{or06} presented polarimetric $UBVRI$ observation data of NGC 6611,
including those of \citet{or00}.  
They identified the presence of nearby dust clouds located on the Local arm 
that have slightly a larger mean $\lambda_\mathrm{max}$\footnote{the wavelength of maximum 
interstellar polarization} than that of the average interstellar medium.  
The mean position angle of the e-vectors was estimate to be $\sim70^{\circ}$. 
However, these optical observations will only be relevant for stars with small extinction and 
hence will not provide information on the magnetic field deep within molecular clouds. 
Thus, infrared polarimetric observations are important.

We conducted deep, near-IR polarimetric observations in M16 to reveal the details of 
M16ES-1/2 and to obtain magnetic field structures in this region.  
We also aimed to search for  more near-IR reflection nebula sources in this region.   
Here we present the results of our  observations and discuss 
the details of M16ES-1/2 and the magnetic field behavior in this region.
The polarimetry instrument used in this study is SIRPOL, which is the attachment of 
the SIRIUS camera mounted on the 1.4-m IRSF telescope at the South Africa Astronomical 
Observatory (SAAO). 
Its distinctive feature is presented in \citet{ka06} and \citet{tam07}.

\section{Observations and Data Reductions}

\begin{figure}
  \begin{center}
   \FigureFile(80mm,80mm){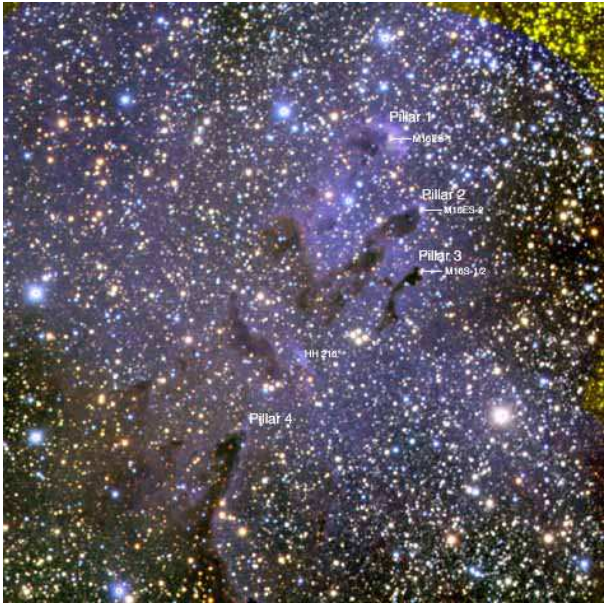}
  \end{center}
  \caption{Three-color composite image of $JHKs$ intensities in M16.
  The $JHKs$ data are represented as blue, green, and red, respectively.
  The area of the image is $\sim$7\farcm7 $\times$ 7\farcm7.
  North is at the top, east to the left. The yellow color regions at the upper right corner 
  and near the middle of the right edge are dead pixel regions of the $J$ band array.}\label{fig1}
\end{figure}

In the southern region of M16 (figure \ref{fig1}), simultaneous $JHKs$ polarimetric
observations were carried out using the SIRIUS camera and its attachment polarimeter
mounted on the 1.4-m IRSF telescope at SAAO.
The SIRIUS camera is equipped with three 1024 $\times$ 1024 HgCdTe (Hawaii) arrays,
$JHKs$ filters, and dichroic mirrors, which enable simultaneous $JHKs$ imaging.
The field of view at each band is $\sim$7\farcm7 $\times$ 7\farcm7, with a pixel
size of 0\farcs45.
The polarimeter is composed of an achromatic (0.5--2.5 \micron) wave-plate rotator
unit and a high-extinction-ratio polarizer, both of which are attached upstream of
the camera and at room temperature.
Details of the SIRIUS camera are presented in \citet{na99} and \citet{na03}, and
those of the polarimeter in \citet{ka06}.

The total exposure time was 900 s per wave-plate angle.
We obtained 10 dithered exposures, each 10 s long, at 4 wave-plate angles
(0$^\circ$, 22.5$^\circ$, 45$^\circ$, and 67.5$^\circ$) as one set of observation.
We repeated this set 9 times.
Sky images were also obtained in between target observations.
The seeing during the observations was 1\farcs1 (2.5 pixel) in the $Ks$ band.

After the standard procedures for near-IR image reduction with IRAF (dark
subtraction, flat-fielding with twilight-flats, bad-pixel substitution, sky
subtraction, and averaging of dithered images), we evaluated the Stokes parameters
($I$, $Q$, $U$), the degree of polarization $P$, and the polarization angle $\theta$
as follows; $Q=I_{0} - I_{45}, U=I_{22.5} - I_{67.5},  I=(I_{0} + I_{45} + I_{22.5}
+ I_{67.5})/2, P=\sqrt{Q^{2} + U^{2}}/I$ and $\theta = (1/2)\arctan (U/Q)$.
The absolute accuracy of position angle of polarization was estimated to be
better than 3$^{\circ}$ based on the measurements during the commissioning run.
The polarization efficiency at $JHKs$ were also estimated to be higher than 96\%
and, therefore, no correction of $P$ was made here.

Aperture polarimetry was performed for point sources detected by DAOFIND in the
field of view.
The DAOPHOT package was used to evaluate the point source magnitudes for 4
wave-plate angles at $H$ and $Ks$. 
The limiting magnitudes (at 0.1 mag error level) of each wave-plate angle are
estimated to be 18.0 and 17.0 at $H$, and $Ks$, respectively.
The 2MASS data were used for magnitude calibration.
All the sources with photometric errors of $\geq$ 0.1 mag were rejected.
The errors of the degree of polarization and position angle were calculated from the
photometric errors, and the degrees of polarization were debiased \citep{wk74}.

\section{Results and Discussion}
\subsection{Intensity and polarization images of the overall region}

\begin{figure}
  \begin{center}
   \FigureFile(80mm,80mm){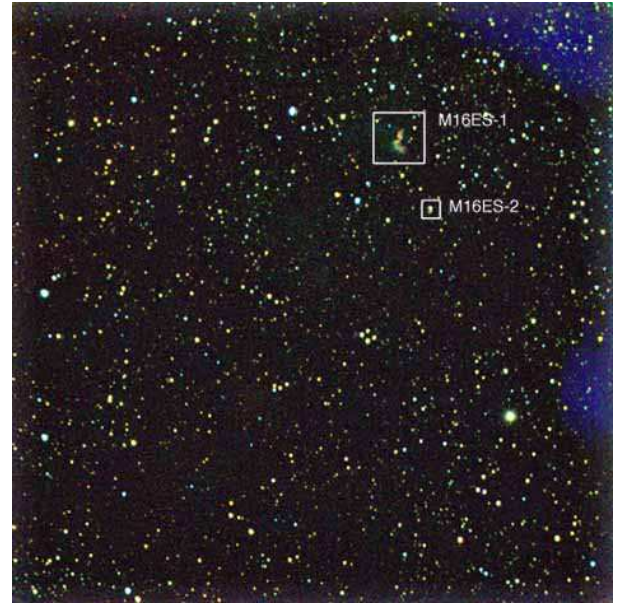}
  \end{center}
  \caption{Three-color composite image of polarized intensity in M16.
  The $JHKs$ data are represented as blue, green, and red, respectively.
  The area of the image is $\sim$7\farcm7 $\times$ 7\farcm7.
  North is at the top, east to the left. The blue color regions at the upper right corner 
  and near the middle of the right edge are dead pixel regions of the $J$ band array.}\label{fig2}
\end{figure}

The $JHKs$ intensities are shown as a composite color image in figure \ref{fig1}.
In the central region of this image, the famous elephant trunks (Pillars 1--3)
are identified as areas of lower stellar density, together with a molecular cloud
that is at the base of the pillars  (near HH 216).
Toward the south-east region of the image, Pillar 4 and its neighboring molecular
clouds \citep{mc01, an04} are also identified.
Relatively strong extended nebular emission is seen around Pillars 1--3, and there is 
rim-brightening emission of the molecular cloud edges (Pillars 1--4, and a cloud near
HH 216) facing the exciting stars that are located to the north-west, but out of this image.
The rim-brightening emission seems to be reflection from the OB star light and/or
from HII gas at the surfaces of the molecular clouds.
The $JHKs$ polarized intensities ($PI$) are presented as a composite color image
in figure \ref{fig2}, where $PI$ is the product of $P$ and $I$.
Distinctive polarization emission nebulae are only identified toward the two YSOs M16ES-1 
and M16ES-2. 
This suggests that most of the extended nebula emission and rim-brightening
emission are not reflected light from the OB stars, but comes from HII gas 
continuum/line emission.

\subsection{Intensity and polarization properties of YSOs at the tips of Pillars 1 and 2}

\subsubsection{M16ES-1}

\begin{figure}
  \begin{center}
     \FigureFile(80mm,*){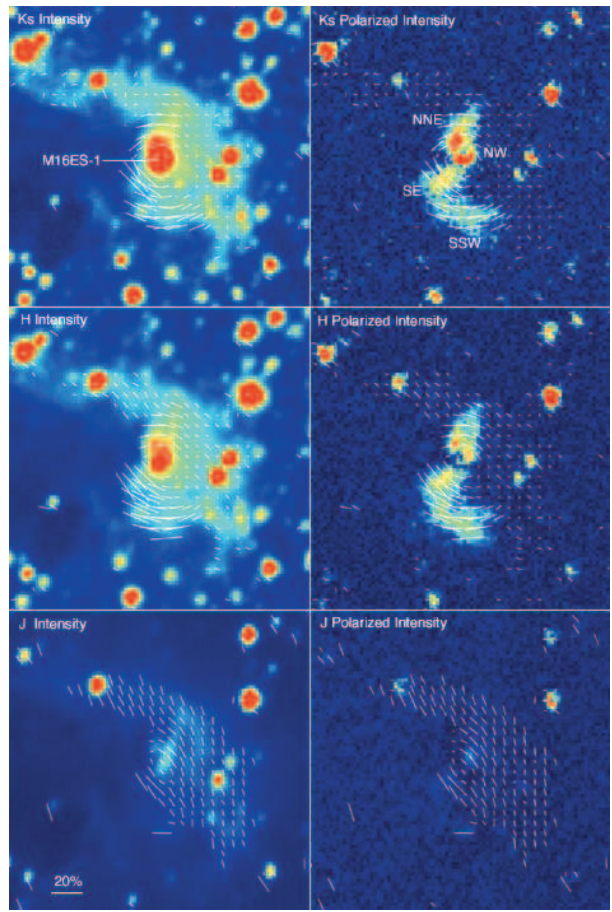}
  \end{center}
  \caption{$JHKs$ polarization vector maps superposed on the total and polarized
  intensity images of M16ES-1 located at the tip of Pillar 1.
  The area of the image is 40\arcsec $\times$ 40\arcsec.
  North is at the top, east to the left. The main exciting star of M16 
  is located to the west-northwest.}\label{fig3}
\end{figure}

A polarization vector map toward M16ES-1 of Pillar 1 is shown superimposed both
on the $I$ and $PI$ images at each band in figure \ref{fig3}.
The appearance of the nebulae in the $PI$ image is significantly different from
that in the $I$ image at each band.
The polarized nebula emission is seen mostly around M16ES-1, while the nebula
emission of the $I$ image is seen both around M16ES-1 and at the rim region facing
toward the exciting stars of M16.
However, the polarization vector maps indicate weak polarized emission in the rim
region, although it is not clearly recognized in the $PI$ images.

Strong polarized emission is asymmetrically distributed just north and south of
M16ES-1 in the $Ks$ and $H$ bands, while point-like weak polarized emission is
seen immediately north of M16ES-1 in the $J$ band.
The polarization vectors of the asymmetric emission clearly show a centrosymmetric
pattern that is probably due to the illumination of M16ES-1, in the $Ks$ and $H$
bands, although this pattern is not clearly seen in the $J$ band.
These suggest that the polarized emission comes from the walls of two (north and
south) cavity lobes, which were created by the bipolar outflow from M16ES-1 and are
still embedded in the molecular gas.
The north lobe has two (NW and NNE) sub-peaks/features and the south one also has
two (SE and SSW) features in  the $PI$ images of $Ks$ and $H$.
The degrees of polarization of the south lobe are 20\% or more, somewhat higher
than those of the north lobe whereas the intensities are higher in the north lobe
than in the south one.

It is possible that these four sub-peaks/features are bright regions within a single bipolar 
cavity lobe, although it is difficult to eliminate the possibility that the outflow system
consists of two pairs of bipolar outflows.
In the latter case, the NW and NNE features would be counterparts of the SE and SSW
features, respectively, suggesting either a binary star system or the precession of
a single bipolar outflow.
\citet{th02} point out the possibility that M16ES-1 consists of multiple low
luminosity objects and our results would appear to support this.
However, the NW and NNE features might be coincident with the two 2.7 mm
continuum peaks y and x of \citet{fu02}, respectively. If they are, then this would 
indicate that the feature are either free-free emission from ultra-compact HII regions, 
i.e., high-mass protostar hypothesis \citep{mc02}, or that they are dust emission from 
very dense regions \citep{fu02}, which are located on top of the cavity with surfaces 
that strongly reflect the light from M16ES-1. 
The polarization vectors toward these two features seem to support the latter possibility.  
The H$_{2}$ image of \citet{th02} shows some emission corresponding to the NW and NNE features,  
which also supports the idea that the two 2.7 mm continuum peaks are dust emission, 
indicating dense cores that have enough masses to form low-mass stars.
Higher resolution observations both at radio and infrared wavelengths, however,
are required to reveal the detailed structures around M16ES-1.

A gap is clearly seen between the NW and SE features (around the center of
the centrosymmetric vector pattern) in the $PI$ images of the $Ks$ and $H$ bands.
This is due to the very-low degree of polarization toward the gap region, since
the intensities at the $Ks$ and $H$ bands are very high there.
This gap most likely corresponds to a small disk-like structure \citep{su02},
perpendicular to the axis of the cavity lobe(s), although the usual ``polarization disk''
(parallel polarization pattern; e.g., \cite{wds87}) is not evident.
This is most likely due to the modest spatial resolution of the observations, 
with a range of polarization position angles from the reflection nebulae 
producing a very small degree of polarization.
If this is the case, a disk-like structure with a size of several thousand AU 
is expected \citep{wh93,fi94}. Here the disk would appear to be tilted 
so that the north side is near to us, producing the higher polarized intensity of 
the north lobe and the higher degree of polarization of the south lobe.

Weak polarized emission is seen along the rim region in the polarization vector
maps of the $H$ and $J$ bands, and the polarization vectors are nearly
perpendicular to the direction of the exciting star(s) from the tip of Pillar 1.
Thus, this polarized emission along the rim region is most likely to be reflected
light from the OB stars of M16, but significantly weaker than that occurs in M42
\citep{tam06}.
The degree of polarization is slightly higher at the $J$ band than at the $H$ band, 
and no alignment of the polarized vectors is recognized at the $Ks$ band.  
These may be due to dilution by line emission (e.g., H$_2$, Br$\gamma$) from HII gas 
at longer wavelengths. 

\subsubsection{M16ES-2}

\begin{figure}
  \begin{center}
    \FigureFile(80mm,*){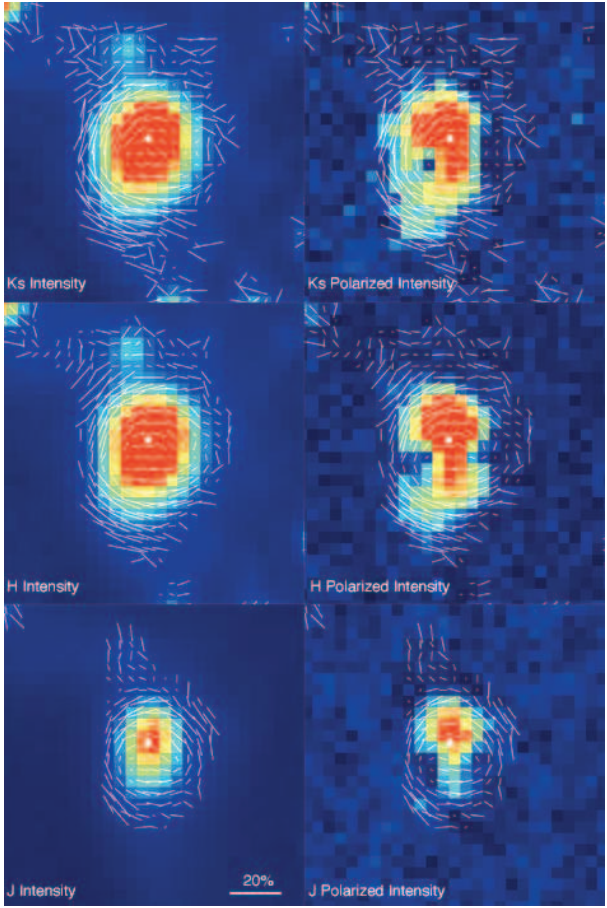}
  \end{center}
  \caption{$JHKs$ polarization vector maps superposed on the total and polarized
  intensity images of M16ES-2 located at the tip of Pillar 2.
  The area of the image is 12\arcsec $\times$ 12\arcsec.
  North is at the top, east to the left. The approximate position 
 of the nebula peak in the HST/NIC 2 image (figure \ref{fig5}) is 
 marked by a white dot at each panel.}\label{fig4}
\end{figure}

A polarization vector map toward M16ES-2 of Pillar 2 is shown superimposed both
on the $I$ and $PI$ images for each band in figure \ref{fig4}.
The polarization vectors toward M16ES-2 clearly show a centrosymmetric pattern,
which is probably due to illumination by M16ES-2, and a polarization disk is seen
at the center of the centrosymmetric pattern.
In the $H$ band, the polarized emission is clearly more constricted 
to the east and west of the approximate center of the centrosymmetric pattern. 
Similar trends are also seen at $J$ and $Ks$.
These, together with the larger total and polarized intensities to the north, 
but larger degrees of polarization to the south (except in the outermost regions) 
are all consistent with a disk-like structure with a size of a few thousand AU,
which is significantly tilted so that the north side is near to us.

\begin{figure}
  \begin{center}
    \FigureFile(80mm,*){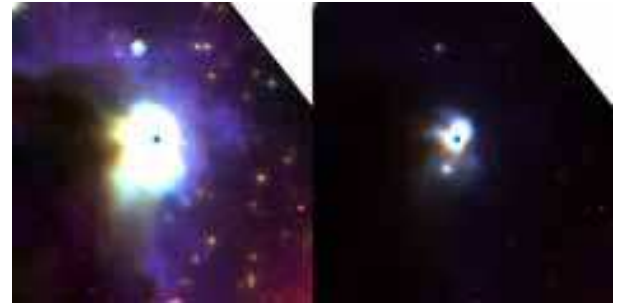}
  \end{center}
  \caption{Three-color composite image of F110W/F160W/F205W intensities
  toward the tip of Pillar 2.
  The F110W/F160W/F205W data obtained with HST/NIC 2 are
  represented as blue, green, and red, respectively.
  North is at the top, east to the left.
  Both images show the same area as that of figure \ref{fig4} with a size of 12\arcsec $\times$ 12\arcsec,
  but the scale of the right one is changed to show
  a constriction of the nebulosity, a faint star, and the difference of intensity between the north and south 
  reflection nebulae.}\label{fig5}
\end{figure}

\citet{su02} reported a dark lane in the middle of the M16ES-2 nebulosity, 
suggestive of an associated disk-like structure, and  
\citet{th02} also suggested that the surrounding nebulosity of M16ES-2 is double.
\citet{mc01} suggested that the double nature of the nebulosity is due to 
a faint second source 1$\arcsec$ south of M16ES-2, which is a lightly reddened field star.
In order to examine further the detailed morphology of the nebulosity, 
we checked the HST/NICMOS camera 2 archival images \citep{th02}.
Figure \ref{fig5} shows the NICMOS 2 infrared image of M16ES-2.
The nebula extent in our total and polarized intensity maps appears consistent
with that in the NICMOS 2 image.   The NICMOS nebula peak is not located at the  
center of the centrosymmetric pattern or the constriction,  but within the north nebula 
(figure \ref{fig4} and \ref{fig5}).
This offset is probably due to the tilt of the disk-like structure and the central source 
might not be seen directly. 
Although a faint source $\sim$1--2$\arcsec$ south of the center (M16ES-2) is identified, 
it is not the origin of the bipolar nebulosity with the intensity of the south nebula 
being much weaker than the north nebula.

We conclude that the bipolar nebula is mainly produced by scattered light from a central 
source obscured by a tilted disk-like structure.

\subsection{Aperture Polarimetry and Magnetic Field Structures}

\subsubsection{Global magnetic field structure}

\begin{figure}
  \begin{center}
\FigureFile(80mm,*){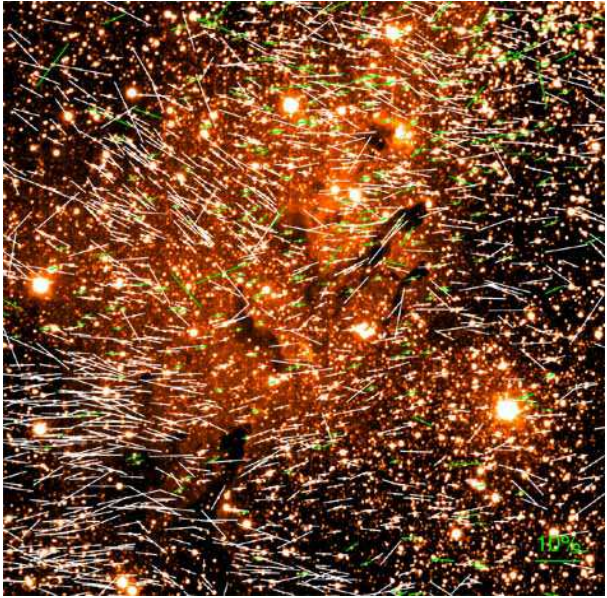}
  \end{center}
  \caption{$H$ band polarization vector map of point sources superposed
  on the $H$ band intensity image.  For sources of $H-Ks$ $\geq$ 0.6 
  their polarization vectors are shown by white bars, for source of 
  $H-Ks$ $<$ 0.6  by green bars.}\label{fig6}
\end{figure}

Figure \ref{fig6} shows the $H$ band polarization vector map of point-like sources
with aperture polarimetry.
Only the sources detected both at $H$ and $Ks$ with photometric errors of $<$ 0.1 mag 
and polarization angle errors of $<$ $15^\circ$ are included.
The bright saturated and crowded sources, which could not be measured
correctly, are rejected.

The polarization vectors are well aligned with each other, and therefore their
polarizations are considered to be produced by dichroism arising from aligned grains 
in the cloud (e.g., \cite{tam87}).
The general regularity of the position angles indicates
a large-scale magnetic field over the observed area, with the polarization vectors 
representing the direction of the magnetic field.

\begin{figure}
  \begin{center}
    \FigureFile(80mm,80mm){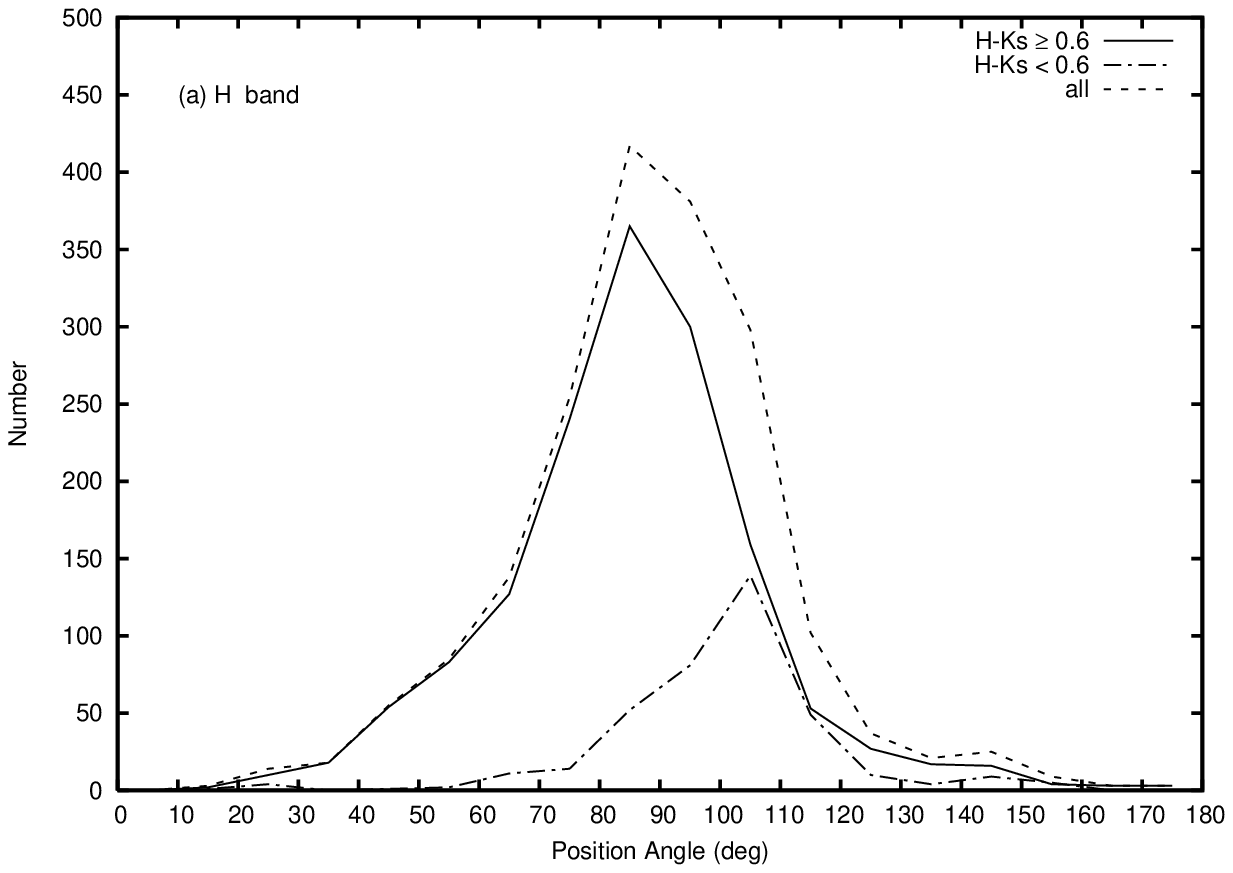}
    \FigureFile(80mm,80mm){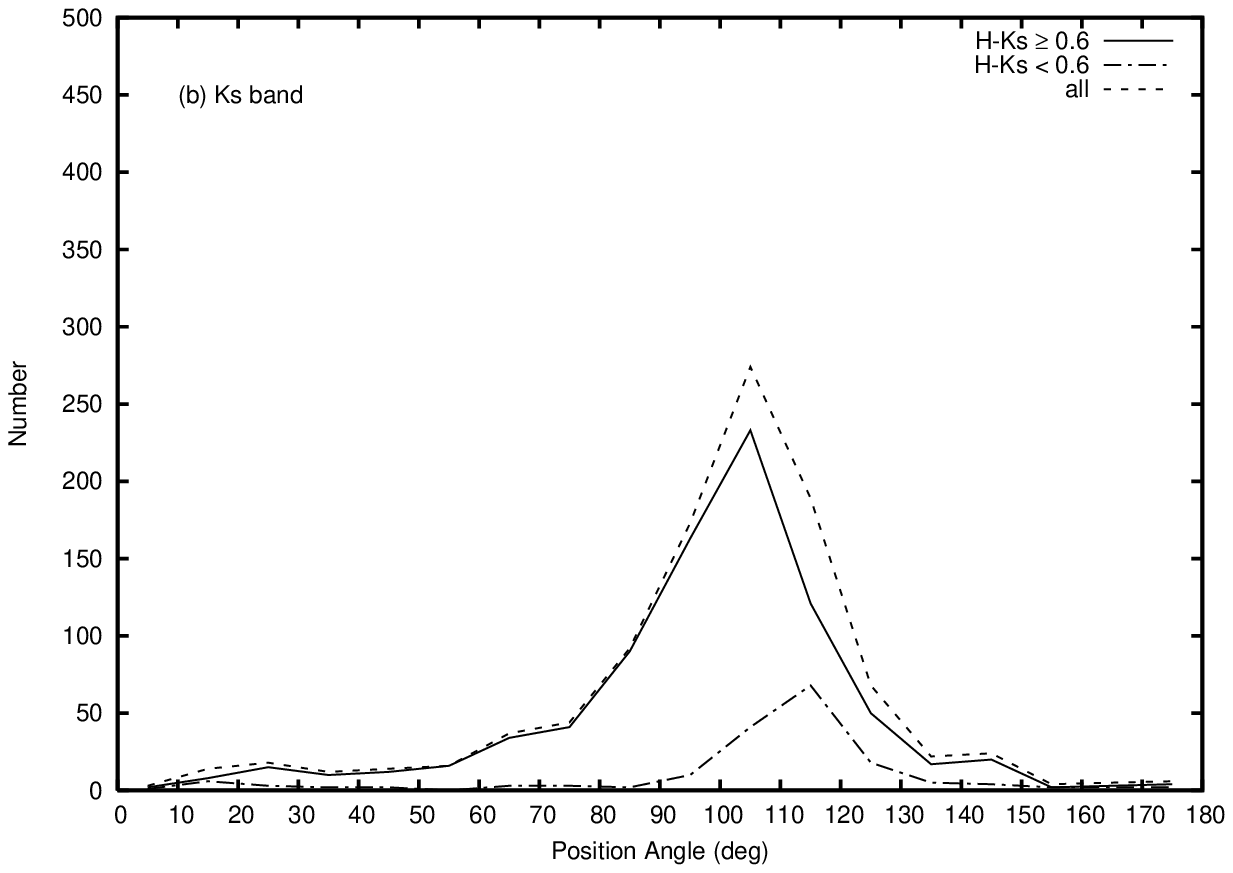}
  \end{center}
  \caption{Histograms of polarization angles of point sources in the $H$ and
  $Ks$ bands. Only sources are included that have photometric errors of $\leq$
  0.1 mag and polarization angle errors of $<$ 15$^\circ$. The width of 
  each bin is 10$^{\circ}$.} \label{fig7}
\end{figure}

The dominant field direction is $\theta$ $\sim$ 80--90$^{\circ}$ with a considerable
spread as shown in figure \ref{fig7} (\textit{dashed lines}).
A small difference in $\theta$ is observed between the $H$ and $Ks$ bands 
[figure \ref{fig7}(a) and figure \ref{fig7}(b), respectively].
The field direction in the north-east quadrant of the observed area is 
$\sim$ 70$^{\circ}$ significantly different from those of the other three quadrants.

\citet{or06} reported $70\fdg0 \pm 2\fdg5$  and $73\fdg7 \pm 2\fdg5$ 
as the mean polarization angles of the stars of the NGC 6611 cluster and its surroundings, 
respectively.
Their values are somewhat smaller than our value of $\sim$ 80--90$^{\circ}$, but 
their optical measurements will only sample stars at lower optical depths.

\begin{figure}
  \begin{center}
    \FigureFile(80mm,80mm){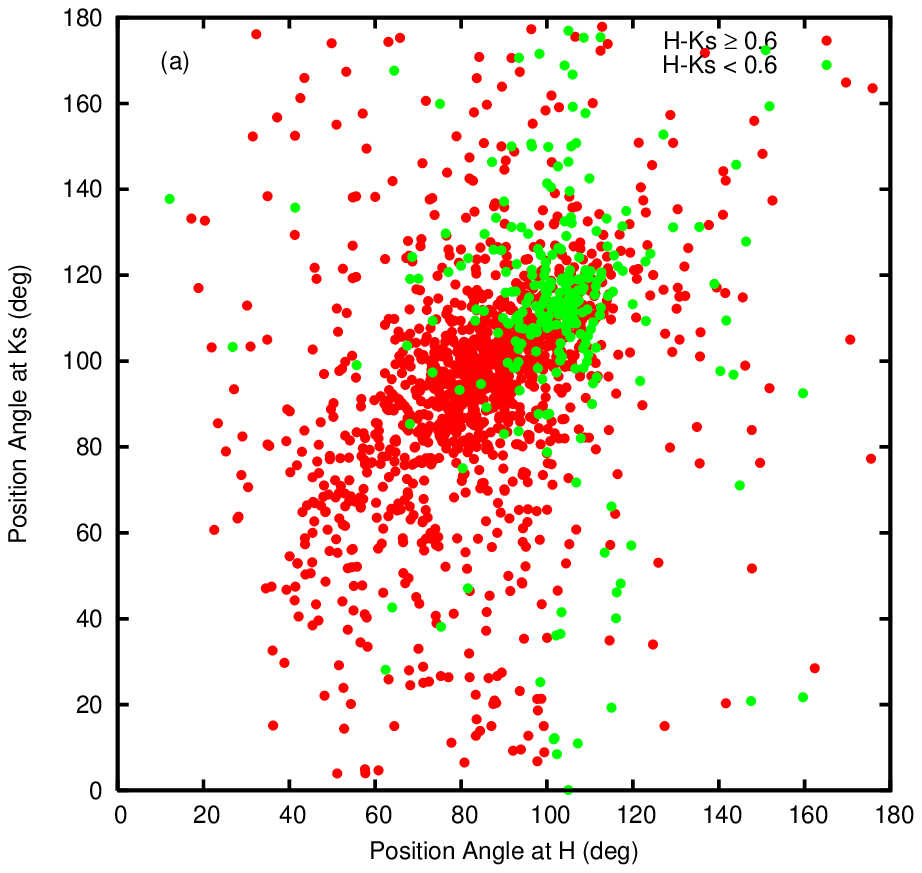}
    \FigureFile(80mm,80mm){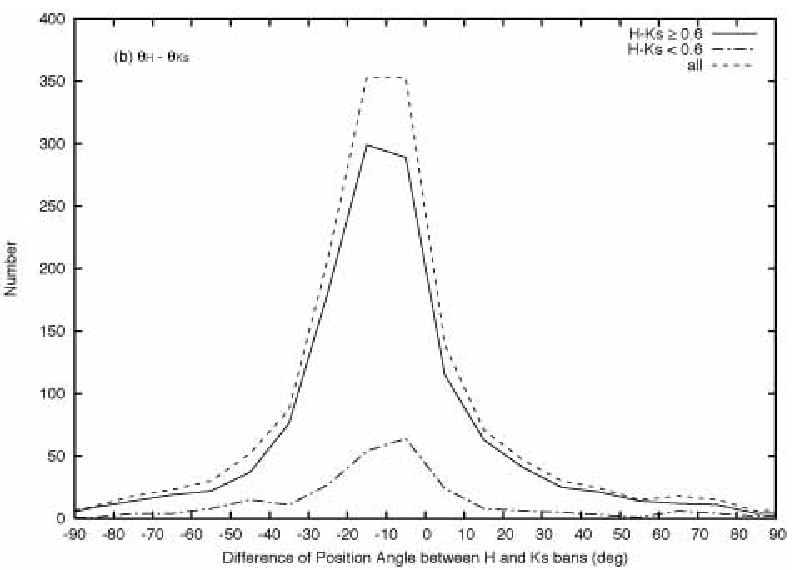}
  \end{center}
  \caption{(a) Correlation of polarization angles between the $H$ and $Ks$
  bands (left) and (b) histogram of the difference of position angles between
  the $H$ and $Ks$ bands (right). 
  Only sources are included that have photometric errors of $\leq$
  0.1 mag and polarization angle errors of $<$ 15$^\circ$. The width of each bin of
  the histogram is 10$^{\circ}$.}\label{fig8}
\end{figure}

To further examine the magnetic field structures in the M16 dark clouds, we divide our sample
into two groups (high and low reddened sources) using $H-Ks$ colors.
According to \citet{su02}, there are two distinct source distributions divided at
around $H-K$ $\sim$ 0.6 in the color-color ($J-H$ vs. $H-K$) diagram.
Many of the highly reddened sources ($H-K$ $\geq$ 0.6) are background stars (dwarfs
and giants) reddened by the M16 clouds, and many of those with low reddening 
($H-K$ $<$ 0.6) are either foreground or stars at low optical depths within the M16 clouds.
The polarization of stars at low reddening may well have a large contribution from interstellar 
polarization, produced in the front of the M16 clouds. The polarization of 
highly reddened stars will be produced mainly within the M16 clouds.
Figure \ref{fig7} shows the histograms of position angles of the observed polarizations for
these strongly and weakly reddened sources (\textit{solid} and \textit{dot-dashed lines},
respectively).
The highly reddened sources have a different mean $\theta$ of $\sim$ 90$^\circ$
($\overline{\theta_H}$ = $84^{\circ} \pm 22^{\circ}$ and
$\overline{\theta_{Ks}}$ = $94^{\circ} \pm 28^{\circ}$) from those with low
reddening ($\overline{\theta_H}$ = $102^{\circ} \pm 18^{\circ}$ and
$\overline{\theta_{Ks}}$ = $110^{\circ} \pm 30^{\circ}$).
The position angles in the $H$ and $Ks$ bands correlate well with each other (figure \ref{fig8}a), 
but differ by $\sim$ 10$^{\circ}$ on the average [figure \ref{fig8}b;
$\overline{\theta_H-\theta_{Ks}}$ = $-10^{\circ} \pm 30^{\circ}$ (for
$H-Ks$ $\geq$ 0.6) and $-8^{\circ} \pm 36^{\circ}$ (for $H-Ks$ $<$ 0.6)].

\citet{ma74} and \citet{me97} interpreted the wavelength dependence of position 
angles to the same sources as arising from 
two clouds along the line of sight, having different dust size distributions 
and different magnetic field directions.
Toward M16, there may exist  at least two types of dust clouds;
one is the clouds associated with M16 and another the interstellar medium
located on the Local arm.
Figure \ref{fig7} shows that the magnetic field direction in the M16 clouds
is $\sim$ 90$^{\circ}$, while $\theta$ of $\sim$ 100--110$^{\circ}$ is suggested in
the interstellar medium.
A tendency for the position angle to be slightly larger at $Ks$ than at $H$ could be 
explained by the larger influence of the interstellar medium at longer wavelengths.
The similar tendency is also seen for sources with smaller $H-Ks$ colors,
probably because a number of sources reddened by the M16 clouds are still
included in the smaller $H-Ks$ color sample.
The optical polarization of sources in \citet{or06} might be mostly dominated 
by the M16 clouds and their $\theta$ might be therefore smaller than ours. 
The larger influence of the interstellar medium at longer wavelengths may
indicate a larger mean dust size in the interstellar medium than in the M16
clouds because a larger dust produces a larger polarization at infrared
wavelengths. 
A larger dust size is consistent with the larger mean $\lambda_\mathrm{max}$ 
in the nearby dust clouds located on the Local arm than in the M16 clouds
that has $\lambda_\mathrm{max}$ similar to the general interstellar medium
\citep{or06}.
SIRPOL observations have shown wavelength dependent $\theta$ 
in other star forming regions and these will be studied in detail in the future.

\subsubsection{Local magnetic field structures}

As shown in figure \ref{fig6}, the north-east quadrant of the observed area has 
$\theta$ $\sim$ 60--70$^{\circ}$, which is  
different from the global value of $\sim$ 80--90$^{\circ}$.   

\begin{figure}
  \begin{center}
    \FigureFile(80mm,*){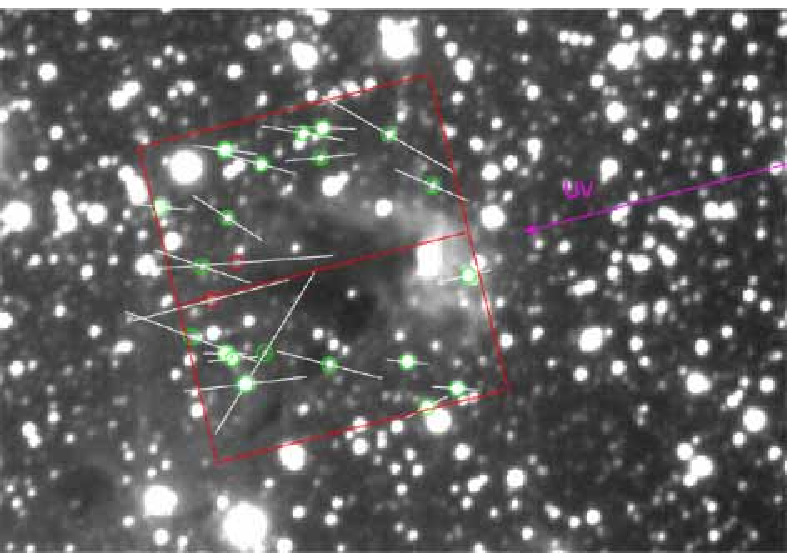}
    \FigureFile(80mm,*){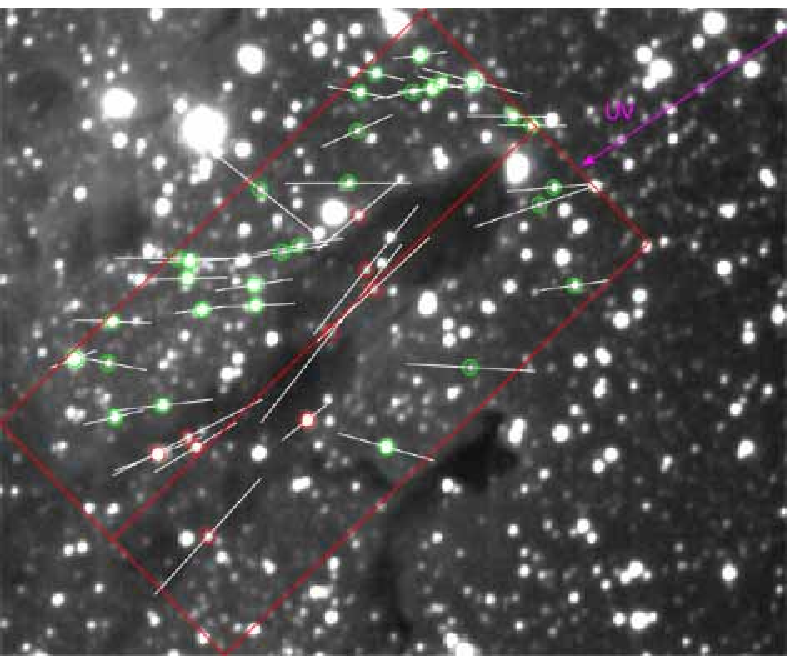}
    \FigureFile(80mm,*){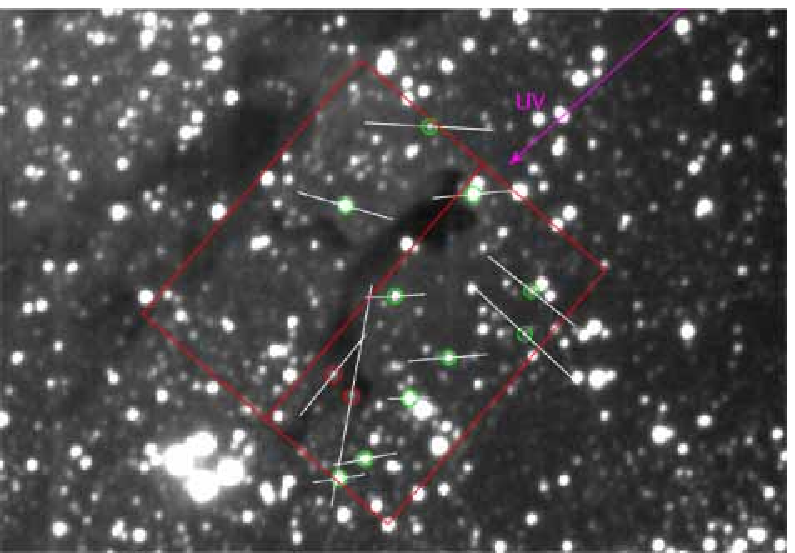}
  \end{center}
  \caption{H band images for position angle estimate in  table \ref{tab}.  
  Sources within molecular clouds 
  are marked by red circles and ones outside by green circles. 
  The width of each image is $\sim$2\farcm5.
  North is at the top, east to the left. 
  See subsection 3.3.2 for more details.}\label{fig9}
\end{figure}

It is interesting to examine the magnetic field direction within the pillars.
The field direction of Pillar 2 seems to be well aligned with the elongation axis of 
Pillar 2, but at quite a different angle to the rest of the cloud.
A similar trend is also seen for Pillar 3. 
In order to examine the relationship in more detail, we list the position angles of 
the elongation axes of the Pillars, the magnetic field directions in and around them, 
and the directions to the Pillars of UV radiation from the main exciting star 
(HD 168076; O5 star) to them.
The position angles of the Pillars were determined by eye, and 
the magnetic field directions by selecting sources within 30$\arcsec$ of their axes, 
which covers most of the length of the pillars (figure \ref{fig9}).   
Only sources with $H-Ks$ $\geq$ 0.6 are included to eliminate foreground 
stars, which for Pillar 1 includ sources only toward its head (see 
figures \ref{fig1} and \ref{fig6}). 

The pillar axes are nearly aligned with the direction of the UV radiation from 
the main exciting star of M16 (table \ref{tab}).
This suggests that Pillars 1--3 might have been strongly affected by the UV radiation   
with their present structures formed through this UV influence \citep{wh99,wi01,fu02,mi06,kb03}.
This strong influence could cause increasing curvature of the clouds' rim,
and the cloud magnetic field could be strongly dragged away from the direction of 
the main exciting star by the shock that precedes the ionization front,  
if the magnetic field is frozen into the gas (e.g., \cite{be89}), 
and then the direction of the field could be eventually aligned with the direction 
toward the exciting star, i.e., the pillar axis direction.   
The  magnetic field direction of Pillar 2 is estimated to be $\sim130^{\circ}$, and nearly 
the same as both the pillar axis and the UV incident direction of $\theta$ $\sim$ 120--130$^{\circ}$, 
while that of the ambient field differs from these $\theta$ by $\sim$ 30--40$^{\circ}$. 
This difference is likely to be the results of the magnetic drag by the ionization/shock front 
in the formation of  Pillar 2.
For Pillar 1 and 3 a similarity is also seen with nearly the same directions 
for the magnetic field, pillar axes and UV radiation, although the numbers of 
polarization angles are very small.  
No clear difference is seen between Pillar 1 and the ambient field directions, 
as might be expected, because the direction of the UV radiation is similar to that of 
the global magnetic field.

The ambient magnetic field directions of Pillars 1--3 are almost the same as that of 
the global field, suggesting the ambient field has not been disturbed.
If these magnetic field directions are those of the initial magnetic field, 
the magnetic field strength could not be large enough to regulate the entire structures 
of the pillars and the ionization gas pressure could be dominant over the magnetic pressure 
in the formation of the pillar structures.
However, we have to wait for a future 3-D hydrodynamical model that fully treats
magnetic effects to know the detailed behavior of magnetic fields in and around the pillar. 

At the tips of Pillar 1--3, signs of recent star formation have been reported and this study 
also suggests that M16ES-1/2 are in the early evolution stage of stellar objects.
The magnetic field directions suggest the pillar structures are strongly influenced by the UV radiation.
Thus, it is very likely that these YSOs were also formed under the influence of the ionization/shock fronts introduced by the UV radiation, although it is also possible that stars were naturally formed 
in dense cores and that then ionization/shock fronts come through the dense cores.
However, YSOs were identified only toward the pillar tips. If stars form naturally 
in the pillars, some YSOs could be identified toward the pillar tails. 
The alignment between the pillar axixes and the UV incident directions imply 
that the ionization/shock fronts made the entire pillar structures from the less dense 
natal clouds and, then, their substructures as shown by hydoromagnetic models 
 (e.g., \cite{be89,le94,wh99, wi01, kb03}).  
Particularly, \citet{kb03} imply that  
a radiatinally impoded globule repeats leaving small cores decupled from itself, which collapse 
after the passage of the ionization/shock fronts and form stars,  i.e., sequential star formation in the globule. 
If this repeated process occur at the pillar tip, the sequential nature of star formation, 
such as mentioned in section 1, could be good evidence for induced star formation. 

Thus, the magnetic fields along the pillar axes could indicate the
strong influence with UV and could consequently provide the possibility of induced star formation 
together with star formation signs at the pillar tips/surfaces.
However, since the sample number is very small at present, it is vitally necessary to 
increase the sample number to further know the general behavior and role of the magnetic 
field in the evolution and star formation of bright-rimmed clouds (pillars around HII regions).

\section{Conclusion}

We have conducted deep and wide ($\sim$7\farcm7 $\times$ 7\farcm7) $JHKs$ imaging
polarimetry in the south region of M16.   Main findings are summarized as follows.

\begin{enumerate}
\item The extended nebular emission around Pillars 1--3 is not reflection light of the OB stars of NGC 6611, but the continuum/line light of the HII region.
\item Strong reflection light can be detected only toward the two nebulae illuminated by 
 M16ES-1 and M16ES-2, at  the tips of Pillars 1 and 2, which appear to be class I objects having circumstellar envelopes.
\item Our polarimetry results implies that M16ES-1 is an intermediate-mass object or lower-mass objects.
\item The mean position angle of global magnetic field is measured to be $\sim$ 80--90$^{\circ}$, 
and seems to be consist with that of the previous optical polarimetry.
\item The alignment among the magnetic fields, axes and UV incident directions of Pillars 1--3, 
and the misalignment of the ambient/global magnetic fields to these directions suggest that the fields are affected by the dynamical impact of the UV radiation.
\end{enumerate}

We thanks T. Nagata and S. Sato for helpful comments.
This research was partly supported by MEXT, Grant-in-Aid Scientific Research on
Priority Area, ``Development of Extra-solar Planetary Science'', and by grants-in-aid from MEXT (Nos. 16077101 and 16077204).  M. T. and R. K. acknowledge support by 
Grant-in-Aid (No. 16340061).

%%%%%%%%%%%%%%%%%%%%%%%%%%%%%%%

\onecolumn
\begin{table}
  \caption{Position angles of the magnetic fields, pillar axes and directions from HD 168076 (O star).}\label{tab}
  \begin{center}
  \begin{tabular}{lcccc}
  
   \hline\hline
                      &           & Magnetic Field &  Pillar Axis & UV Incident Direction \\
      Region && (deg) & (deg) & (deg) \\
    \hline
      Pillar 1 & inside & 99$\pm$5 (2) & $\sim$104 &   $\sim$104 \\
                    & ambient & 87$\pm$21 (19) &  ... & ... \\
      Pillar 2 & inside & 129$\pm$10 (9) &  $\sim$134 &  $\sim$123 \\
                    & ambient & 91$\pm$13 (30) & ... & ... \\
      Pillar 3 & inside & 155$\pm$14 (2) &   $\sim$139 &  $\sim$132 \\
                   & ambient & 83$\pm$18 (10) & ... & .... \\
      Global & ... & 84$\pm$22 (1227) & ... & ...  \\
    \hline 
    \multicolumn{5}{@{}l@{}}{\hbox to 0pt{\parbox{85mm}{\footnotesize
      Note.-- The position angles of magnetic fields were derived from $H$ band data,
      and numbers of sample are given in parentheses.
      }\hss}}
    \end{tabular}
  \end{center}
\end{table}

\end{document}